%% file: main.tex
\documentclass[format=acmsmall, review=false, screen=true]{acmart}

\usepackage{booktabs} 
\usepackage[ruled]{algorithm2e} 

\SetAlFnt{\small}
\SetAlCapFnt{\small}
\SetAlCapNameFnt{\small}
\SetAlCapHSkip{0pt}
\IncMargin{-\parindent}
\renewcommand{\footnotesize}{\tiny}

\usepackage{amsmath}
\usepackage{booktabs} 
\usepackage[shortlabels]{enumitem}
\usepackage{geometry}
\usepackage{graphicx}
\usepackage{hyperref}
\usepackage{IEEEtrantools}
\usepackage{import}
\usepackage[utf8]{inputenc}
\usepackage{mathtools}
\usepackage{multirow}
\usepackage{subcaption}
\usepackage{url}

\graphicspath{{./images/}}

\acmJournal{CSUR}
\acmVolume{1}
\acmNumber{1}
\acmArticle{1}
\acmYear{UNDER REVIEW}
\acmMonth{0}
\copyrightyear{UNDER REVIEW}

\setcopyright{acmlicensed}

\acmDOI{0000001.0000001}

\begin{document}
\title[Case law retrieval]{Case law retrieval: accomplishments, problems, methods and evaluations in the past 30 years}
\author{Daniel Locke}
\affiliation{%
  \institution{The University of Queensland}
  \city{St Lucia}
  \state{QLD}
  \postcode{4067}
  \country{AUS}}

\author{Guido Zuccon}
\affiliation{%
\institution{The University of Queensland}
\city{St Lucia}
\state{QLD}
\postcode{4067}
\country{AUS}}
\input{abstract}
%
%

%
%



\maketitle

\input{introduction}
\input{case-description}
\input{literature-review.tex}

\input{open-issues}
\input{conclusion}

\vspace{-5pt}

\bibliographystyle{ACM-Reference-Format}
\bibliography{bibliography}

\end{document}

%% file: abstract.tex
\begin{abstract}
Case law retrieval is the retrieval of judicial decisions relevant to a legal question. Case law retrieval comprises a significant amount of a lawyer's time, and is important to ensure accurate advice and reduce workload. We survey methods for case law retrieval from the past 20 years and outline the problems and challenges facing evaluation of case law retrieval systems going forward. Limited published work has focused on improving ranking in ad-hoc case law retrieval. But there has been significant work in other areas of case law retrieval, and legal information retrieval generally. This is likely due to legal search providers being unwilling to give up the secrets of their success to competitors. Most evaluations of case law retrieval have been undertaken on small collections and focus on related tasks such as question-answer systems or recommender systems. Work has not focused on Cranfield style evaluations and baselines of methods for case law retrieval on publicly available test collections are not present. This presents a major challenge going forward. But there are reasons to question the extent of this problem, at least in a commercial setting. Without test collections to baseline approaches it cannot be known whether methods are promising. Works by commercial legal search providers show the effectiveness of natural language systems as well as query expansion for case law retrieval. Machine learning is being applied to more and more legal search tasks, and undoubtedly this represents the future of case law retrieval. 
\end{abstract}

%% file: introduction.tex
\section{Introduction}
Law as a profession comprises a significant academic basis in that it focuses on finding, reading and analysing existing laws in various forms so as to advise a person on the legal bearing of their situation. Case law is one of the main sources of law in common law jurisdictions. This survey focuses on the task of finding relevant case law.

The focus of a lawyer's job on analysing case law comes about from the principle of \textit{stare decisis} (doctrine of precedent). So that lawyers can advise clients, they must be able to find these binding legal principles.\footnote{~\textit{Stare decisis} "requires, broadly speaking, that like circumstances are considered in a like fashion; a case that considers a certain set of factual circumstances ... [is precedential] for any [analogous] future circumstances" and must be followed~\cite{AIRS}. } The need and importance for lawyers to find these cases is so "[they] can discharge their duty owed to the court and to the administration of justice"~\cite{AIRS}. Finding case law accounts for roughly 15 hours per week for a lawyer~\cite{Lastres_2013} or nearly 30\% of their yearly working hours~\cite{Poje_2014}.\footnote{~This figure was for junior lawyers, with less than 10 years experience. Lawyers with more than 20 years experience spent on average 16\% of their time researching. This may be due to experience or a change in role to a more managerial position. While more senior lawyers spend less time researching, their work is typically at a much higher cost to a client. } 

Within the context of finding case law, the amount of information that must be searched is large, and it is growing rapidly. In the US, by 1962, 25,000 opinions were being published each year, and there were over 2.3 million decisions in the published case law reports~\cite{Wilson_1962}. By the early 2000s, this number had increased significantly; US appellate courts alone were handing down about 500 decisions each day~\cite{Jackson_2003}. Other legal sources also include large amounts of data: modern search providers search over a billion documents~\cite{BLAWIndex}.

Throughout the last 20 years various methods have been identified as the standard for case law retrieval. In 2001, Moens~\cite{Moens_2001} stated some form of concept based retrieval, relying on manual indexing of terms remained commonplace for case law retrieval. However, with the large number of cases being published, manual indexing was becoming ever more less feasible. She identified three models as standard for case law retrieval: Boolean retrieval, vector-space retrieval, and probabilistic retrieval. Maxwell and Schafer~\cite{Maxwell_2008} identified similar models as the standard for case law retrieval in 2008, in addition to noting that knowledge engineering based methods for retrieval are common. Today, separate retrieval and ranking phases are the standard~\cite{ConradNextGen2013}. These systems score documents using multiple weighted features such as the court, user jurisdiction, citations and past query logs~\cite{Liao2009, ConradNextGen2013, Pompili2019}. Likewise, use of deep learning techniques in various stages of the retrieval process is becoming more common place, as we discuss throughout.

It has been over 20 years since Turtle~\cite{Turtle_1995} and nearly 20 since Moens~\cite{Moens_2001} surveyed legal information retrieval. Yet in this time, little research has focused on ranking for case law retrieval as compared to information retrieval generally. Published research as to ranking for ad-hoc case law retrieval lags behind the general information retrieval community, despite legal information retrieval being one of the first applications of computers to information retrieval~\cite{Bing_2010}. This is likely an incident of commercial search providers choosing not to publish any research. Aside from Moens~\cite{Moens_2001} and Maxwell and Schafer's~\cite{Maxwell_2008} identification of the standard methods for case law retrieval, there is only limited research on improving the effectiveness of these methods, or \textit{a priori}, identifying or comparing their effectiveness. The majority of research in the legal information retrieval field has focused on argumentation retrieval, ontological frameworks and case-based reasoning, and technology-assisted review for discovery~\cite{Baron2006, Cormack2010}.\footnote{~Discovery is the process by which parties to a dispute disclose all documents relevant to the issues between them (see rules 211 and 212 of the \textit{Uniform Civil Procedure Rules 1999} (Qld).)}

This survey focuses on the task of querying for ad-hoc\footnote{~Ad-hoc refers to the task of generic querying to satisfy an information need~\cite{Croft2015}. } case law retrieval. But, legal information retrieval does not end there. Lawyers perform many search tasks. For instance, lawyers may search for legislation (laws enacted by parliament) (this may be ad-hoc search or by providing a factual scenario similar to ontological frameworks~\cite{Wang2018, Wang2019}), civil codes (similar to legislation in a civil law jurisdiction)~\cite{Heo_2017, Kim_2017, Kim_2014, Kim_2016}, for documents in litigation, such as technology-assisted-review~\cite{Baron2006, Cormack2010}, for patents, company information, dockets, or other documents generally in internal support system of a law firm~\cite{Moens_2001}. Vast amounts of work have been done in these areas (for instance the TREC Legal tracks~\cite{Baron2006, Cormack2010} and ICAIL DESI series~\cite{Baron2007, Baron2015} ). Despite the existence of these other tasks, case law retrieval remains the sole focus of this article because of the unique challenges research in this area has encountered. Searching for case law is a peculiarity of common-law jurisdictions; the same need to find cases does not exist in a civil law jurisdiction.

By this article, a survey of work to date in the field of case law retrieval is examined. The article continues as follows: in Section~\ref{sec:case-desc} we outline the parts of a case; in Section~\ref{sec:history} we describe in detail the nature of case law retrieval and provide a history of initial research in the area; in Section~\ref{sec:collections} we outline Cranfield style collections used in the evaluation of methods for case law retrieval; and, in Section~\ref{sec:methods} we outline various methods for case law retrieval as well as gaps in the literature and promising areas for future research.

%% file: case-description.tex
\section{Overview of a case}
\label{sec:case-desc}

\begin{figure}[t!]
\includegraphics[width=0.4\linewidth]{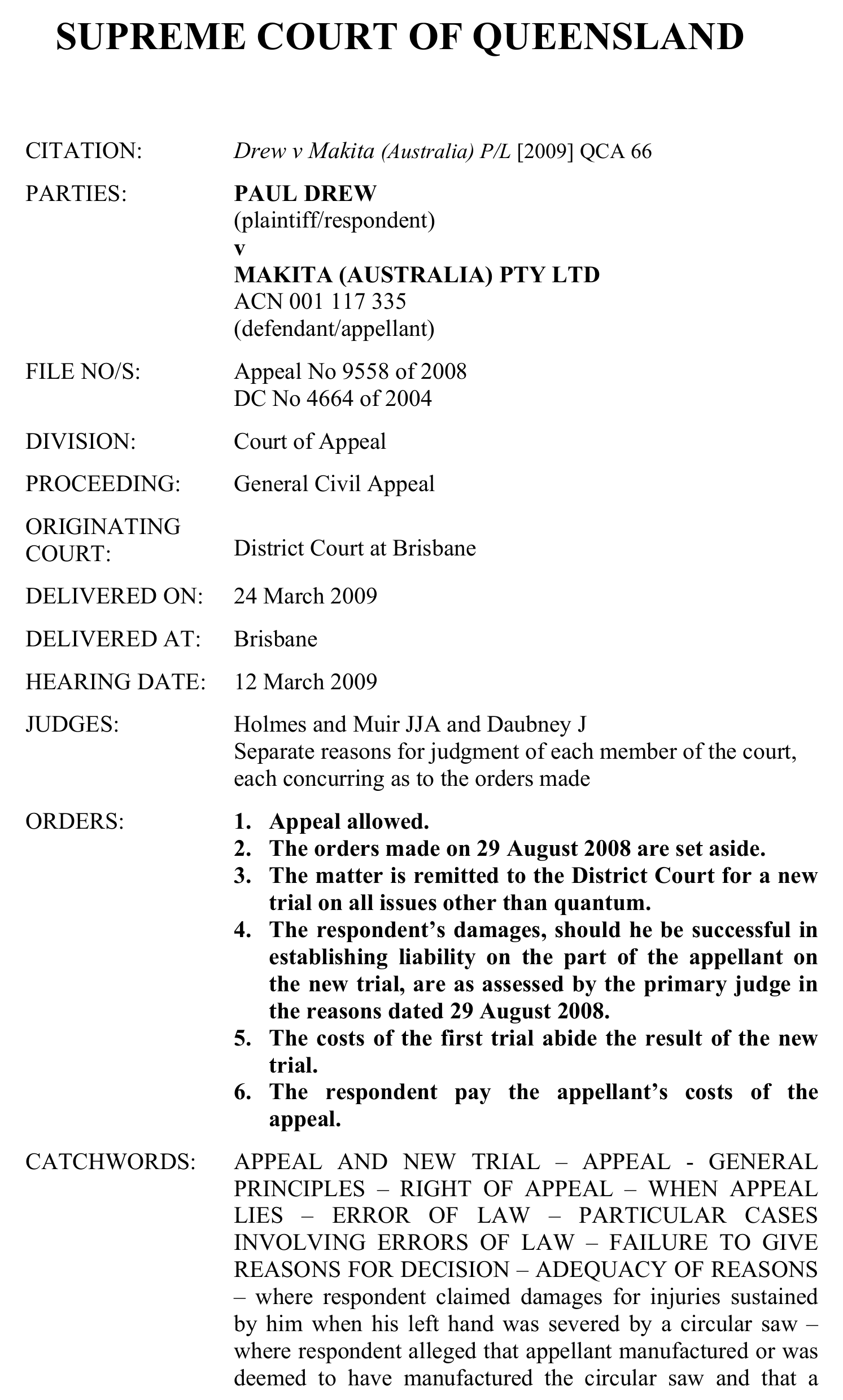}
\vspace{-10pt}
\caption{Portion of a headnote of a decision of the Queensland Court of Appeal: \textit{Drew v Makita (Australia) Pty Ltd} [2009] QCA 66.}
\label{fig:headnote}
\end{figure}

\begin{figure}[t!]
\vspace{-15pt}
\includegraphics[width=0.8\linewidth]{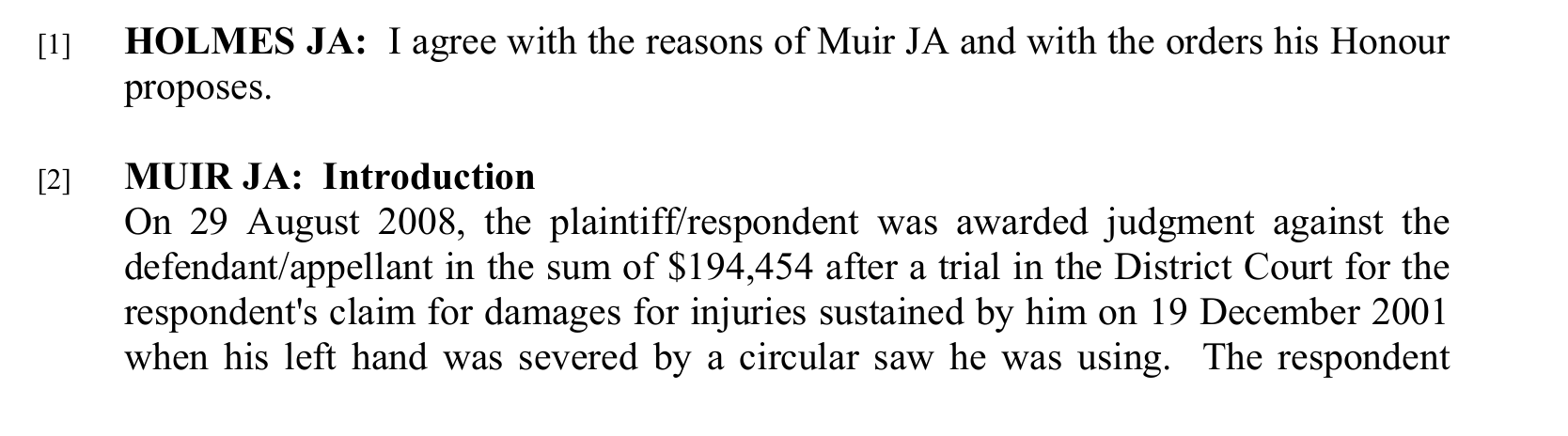}
\vspace{-15pt}
\caption{The beginning of the reasons for decision of \textit{Drew v Makita (Australia) Pty Ltd} [2009] QCA 66 \vspace{-10pt}}
\label{fig:judgment}
\end{figure}

In Figure~\ref{fig:headnote} we show a headnote of an example case, and in Figure~\ref{fig:judgment} we show the start of the text of the same case's reasons. The headnote is not part of the decision and contains only metadata about the case.~\footnote{This differs to an American lawyers' understanding of a headnote, which refers to editorial annotations that summarise one point of law in a case. } Nonetheless, information in the headnote is relevant for filtering and searching on cases. The citation, and file numbers (FILE NO/S) may be searched on in the case of known-item retrieval. Likewise, the division, proceeding, delivery date, or judges may all be used for filtering results given an ad-hoc search. Typically the catchwords would be the only field outside of the text of the judgment that will be searched on. These catchwords are manually identified descriptions of the case provided by judges or court staff, and will typically include a list of cited cases and  legislation. 


%% file: literature-review.tex
\input{literature-review/history.tex}
\input{literature-review/collection.tex}
\input{literature-review/methods.tex}

\input{literature-review/network-analysis.tex}
\input{literature-review/rhetorical-roles.tex}

\input{literature-review/query-expansion.tex}

%% file: literature-review/history.tex
\section{The task of case law retrieval, and an history}
\label{sec:history}
Within the task of case law search, the following tasks can be identified: (i) general ad-hoc querying; (ii) question answering; (iii) routing or selective dissemination; (iv) known item search; (v) navigation; and (vi) intra-document retrieval~\cite{Turtle_1994}. Querying has been said to pose the greatest problem for lawyers, in that they must: ``(1) know and be able to articulate their information need; (2) know the content and storage structure of the documents in the database; and (3) be aware of the operation of the retrieval mechanism.''~\cite{Matthijssen_1998}. While this survey focuses on the task of ad-hoc querying for case law retrieval, there will necessarily be overlap between the applicability of methods and concepts and a discussion of their results with the other tasks listed above.

In 2010, Bing~\cite{Bing_2010} documented the history of case law retrieval from a practical perspective. This was less focused on any particular commercial search provider's system but with the early genesis of case law retrieval. In 1995, Turtle~\cite{Turtle_1995}, on the other hand, provided more of an academic view of the history to date of case law retrieval. Given Turtle's work at Westlaw, his work is more detailed as regards their methods. Apart from the short summary below, we do not endeavour to repeat either of their syntheses. Rather, we focus on the developments made in the last 20 years. Turtle highlights the rapid developments of case law retrieval from the 1970s through to the 1990s. He states that prior to the 1970s ``virtually all legal research was based on printed materials.'' At that point in time, major electronic providers only provided first-order logic based retrieval. But, by the early 1990s ordinary language systems had begun to emerge. Turtle~\cite{Turtle_1995} and Moens~\cite{Moens_2001} discussed a combination of the following points of case law retrieval, which we further elaborate on:
\vspace{-4pt}
\begin{enumerate}[label=\textbf{(\arabic*)}, itemindent=0em, labelsep=0.5em,leftmargin=*]
	\item \textit{the collections} that must be searched on are large (at the time of Turtle's~\cite{Turtle_1995} review, US legal text comprised about 50 gigabytes of data, and was growing at a rate of 2 gigabytes per year). A subset of all legal documents, being those that are publicly available in the US consists of approximately 240 gigabytes of data. This comprises roughly 4 million documents. By comparison, in 2012, commercial systems were searching on over 11 million cases~\cite{Mart_2013};
	\item \textit{the document length} is greater than those in other areas such as web search, but also varies widely. As an extreme example, the decision of \textit{Bell Group Ltd (in liq) v Westpac Banking Corporation} (2008) 39 WAR 1 is 961 pages long, and at the other end of the spectrum, the decision of \textit{Western Export Services Inc v Jireh International Pty Ltd} (2011) 282 ALR 604 is 6 paragraphs long. One must add that in the context of ad-hoc case law retrieval, the full text, or the catchwords, as a summary of the case, are the only fields that will frequently be searched on. The title is not descriptive of the issue discussed, and represents only the parties involved in a dispute; as such, the title is only relevant for known item retrieval. The full text rather than a summary or metadata is the most important aspect~\cite{van_Opijnen_Santos}. Further, not all documents will have a headnote summarising the issues. In this respect, headnotes are expert compiled either by an editor or by the respective decision maker and may be of varying quality. But, searching on headnotes represents one aspect in which search may progress. More modern controls over the publishing of judicial decisions has led to greater availability of headnotes.
	\item \textit{there is significant added editorial value} with indexing, summaries, notes, and classification codes. Commercial services add summaries of decisions which may also be searched on. They may also classify decisions according to their topic or topics thereby aiding in retrieval (for example, Westlaw Key numbers\footnote{~See \url{legal.thomsonreuters.com/en/insights/articles/using-the-west-key-numbers-system}});
	\item \textit{the relationship of the institution} creating the legal document is important (which court makes the decision has a bearing on its importance). Such an approach has been considered in early citation analysis approaches to find decisions relevant to a seed decision~\cite{Tapper_1981}. Tapper considered the differing jurisdiction of a decision in seeking to give a numeric value to a citation for the purpose of clustering. This sought to give an inverse value to a decision based on the level at which it was decided, with a higher court at the top, such as the Supreme Court of the United States, or the High Court of Australia receiving a lower value, and a lower court, such as an intermediate appellate court receiving a higher value. While Turtle does not make the point, so too may the specific author be important. Westlaw accords some weight to the court and user's jurisdiction in ranking case law~\cite{ConradNextGen2013}. For instance, Ravel Law\footnote{~A recently created commercial search engine:~\url{ravellaw.com}} allows searchers to explore the specific decisions of a particular judge to analyse~\cite{Ravel_judges}. In fact, it premises a large number of its features on this ability.
	\item \textit{legal databases are the law} and while in other fields, old documents may not reflect current state of doing things (such as in evidence based medicine), cases from 200 or 400 years prior may still be applicable or informative;
	\item \textit{the language} is unique. Judicial language has been said to be rich and feature very domain specific terminology~\cite{van_Opijnen_Santos}, to feature no redundancy~\cite{Schweighofer_Geist} and that it is not a pure sublanguage~\cite{Moens_2001}. One thing is certain: legal sentences are frequently long and contain many clauses;
	\item \textit{citations} are very important in legal decisions. The doctrine of precedent requires that courts that sit lower in the hierarchy must follow the decisions of higher courts. Also, judges must explain their reasons for a decision. To show that judges are conforming to binding authority they can explain their approach and explain how it accords within previous decisions. Hence, there is a rich network of citations. But, their effectiveness in aiding retrieval is not explored;
	\item \textit{the audience} is wide, and ranges from laypersons to academics and lawyers~\cite{van_Opijnen_Santos};
	\item \textit{the cost} of legal research is high. Subscriptions to commercial systems cost individuals thousands of dollars each year. Lawyers charge fees for their services, usually based on the time taken. Accordingly, minimising the time taken researching a legal question is key to reducing end costs for clients and making the legal system more accessible and affordable. 
\end{enumerate}

\subsection{What is the nature of case law retrieval?}
The nature of case law retrieval is an important question. It determines whether methods or systems for retrieval are effective. Maxwell and Schafer~\cite{Maxwell_2008} summarised the early positions of Dabney~\cite{Dabney_1986, Dabney_1993} that total recall is required. This position is understandable. It takes the view that as lawyers are responsible for their work they must be fully informed otherwise they face adverse cost awards or ethical breaches. This is the view taken by Lu and Conrad~\cite{LuConrad2012}. Such a position is common to all professional tasks (e.g. technology assisted review)~\cite{Tait2014, RussellRose2018}. But legal work is frequently cost and time constrained. Gerson~\cite{Gerson_1999} suggested, however, that the task is merely one that is precision orientated. This is because of the amount of decisions a lawyer will be willing to read within a reasonable time and budget, and because of the existence of other tools such as those that allow a lawyer to follow the citation history of a case. He stated that because of the nature of legal information, where cases are processed and summarised by legal publishers, the key to legal research is to find one relevant decision quickly, from which other tools can be used to find other relevant decisions. Therefore, a legal information system should be judged by how quickly it can find relevant decisions~\cite{Gerson_1999}. This is certainly an attractive approach to legal research and it is one where tools, such as Zhang and Koppaka~\cite{Zhang_Koppaka}, proposed and those enabling searchers to trace citations by paragraph,\footnote{~See for example \url{jade.io}.} so they can focus on the distinct legal question relevant, gain use.

Gerson~\cite{Gerson_1999} shared the view that the task is a recall orientated task. He places an important proviso on this view: that it is also one that requires maximisation of precision. The maximisation of precision is necessary to: (i) ensure that lawyers are not overburdened with information; and (ii) reduce costs associated with legal research and the provision of legal services. It is this view that the authors share. This view is sensible in light of the costs of legal services, whilst ensuring that quality is maintained. In this sense, case law retrieval is different to total recall tasks such as prior-art search in patent retrieval~\cite{Shalaby2019} or systematic review literature searching~\cite{Higgins2011}. Lu and Conrad~\cite{LuConrad2012} appear to share a similar view of case law search as a high recall task that is cost constrained. Case law may, more appropriately, be considered interactive.\footnote{~E-discovery may also be considered highly interactive, with human in the loop processes~\cite{GrossmanCormack2010}. } This has been long recognised~\cite{Harrington_1984}. Turtle~\cite{Turtle_1994} for instance, identified case law retrieval as an interactive task, with searchers willing to only look at somewhere around the top 20 results. The authors share this view. Recent discussion of queries used in commercial case law retrieval systems may support this~\cite{Shankar2018}.


%% file: literature-review/collection.tex
\section{Available collections for case law retrieval}
\label{sec:collections}
Test collection based evaluation of information retrieval has represented the standard since the 1950s~\cite{Sanderson_2010, Voorhees2005}. Despite this, until recently, there has been no standard collection of general utility for evaluating case law retrieval. And in this regard, work is still required. In Table~\ref{tab:collections}, we outline current available collections for case law retrieval.

\begin{table}[t]
\centering
\begin{tabular}{rrrrr}
\hline
\textbf{Collection} & Documents & Queries & Assessments & Purpose \\
\hline
Locke and Zuccon~\cite{SigirCollection} & 3,597,230 & 12 & 2572 & general use \\
Locke et al.~\cite{AIRS} & 63,916 & 100 & 2645 & query generation \\
Koniaris et al.~\cite{Koniaris_Multi_Dimension} & 63,742 & 330 & 105,036 & diversification \\
Koniaris et al.~\cite{Koniaris_Eval_Diversification} & 3,890 & 330 & unknown & diversification \\
\end{tabular}
\caption{Collections previously created for use in case law retrieval. \vspace{-15pt}}
\label{tab:collections}
\end{table}

Locke and Zuccon~\cite{SigirCollection} summarized the collections available for case law retrieval. Aside the 2017 collection of Koniaris et al.~\cite{Koniaris_Eval_Diversification} which contains under 4,000 documents, Locke et al.~\cite{AIRS} and Koniaris et al.~\cite{Koniaris_Multi_Dimension} both created collections based on 60,000 United States Supreme Court cases. The collections of Locke et al.~\cite{AIRS} and Koniaris et al.~\cite{Koniaris_Multi_Dimension} contain the same documents.

Locke and Zuccon~\cite{SigirCollection} have previously said that these collections (and the smaller collection of Koniaris et al.~\cite{Koniaris_Eval_Diversification} to a greater extent) are not representative of how many cases a lawyer may realistically search on. This is because the current total number of legal decisions in common publicly accessible legal search systems range from about 850 thousand cases (Courtlistener) to over 4 million (Austlii). Comparatively, Westlaw\footnote{~\url{www.westlaw.com}, A popular commercial case law search system.} in 2012 reportedly contained 11 million documents that could be searched on~\cite{Mart_2013} and Bloomberg Law presently contains around 1 billion documents~\cite{BLAWIndex}.\footnote{~It is not known what percentage of these are cases as opposed to other document types.} Searching over such a small number of cases may be representative of searching a singular court or jurisdiction's decisions. 

Further, as Locke and Zuccon~\cite{SigirCollection} identified, the collections created by Locke et al.~\cite{AIRS} (AIRS) and that of Koniaris et al.~\cite{Koniaris_Multi_Dimension} are not appropriate for generally evaluating case law retrieval in so far as the assessments are concerned. The collection of Locke et al.~\cite{AIRS} has an average of 26 assessments per topic. While, as Locke and Zuccon noted~\cite{SigirCollection}, a small number of relevance judgments is not appropriate for case law retrieval, as a recall orientated task~\cite{Gerson_1999}, this was not a limitation of their work because they pooled methods prior to assessment. It does mean, however, that the collection cannot be reliably used to evaluate methods that were not pooled. The collection of Locke and Zuccon~\cite{SigirCollection} addresses this issue. However, its small number of queries is problematic. Their collection, which comprises more than 3.5 million documents, contains a minimum of 200 assessments per topic, pooled through a number of standard queries, and expert user issued queries. The methods for the creation of this collection follow similar methods used to create internal collections at Westlaw~\cite{Turtle_1994}.

Koniaris et al.'s~\cite{Koniaris_Multi_Dimension} collection does not have a small amount of assessments. However, the assessments automatically gathered (they did not obtain expert relevance assessments). They took the highest ranked documents from an LDA topic model and labelled these documents as relevant. Looking at the topics used by Koniaris to generate the assessments also highlights the lack of general utility of the collection. The queries are a subset of the areas of law in Westlaw's Digest.\footnote{~A keynumber system of categorised areas and subareas of law. Areas of law can be searched or browsed by number.} They represent a general search need rather than a specific information need. This is distinct from the problem that queries commonly issued in Lexis are very short and have been said to be insufficient to describe an information need~\cite{Shankar2018}. In Table~\ref{tab:queries}, we extract Locke and Zuccon's~\cite{SigirCollection} summary of the queries used by Koniaris et al.~\cite{Koniaris_Multi_Dimension} are compared with those of Turtle~\cite{Turtle_1994}, which represent queries created by an expert searcher (a lawyer), rather than extracted from the Westlaw Digest, as Koniaris et al.~\cite{Koniaris_Multi_Dimension}. This shows the artificial nature and the broad scope of the queries from Koniaris et al.. The queries employed by Koniaris et al.~\cite{Koniaris_Multi_Dimension} may be more reflective of a general search to gain background information, rather than to search for a specific legal issue. 

It is important to note that assessments for case law search collections are difficult to obtain. Relevance assessments based on citations are not appropriate. This is because a decision may cite another decision for any number of reasons, for any number of different topics. Tapper~\cite{Tapper_1974} views objective relevance assessments in the context of Cranfield based evaluation of retrieval systems as preferable: that being to take the citations in a case as relevance assessments rather than asking searchers for subjective assessments. But, he articulates two reasons why this is not appropriate however. First, that the citations may not be adequate. And second, that the citations may not be exhaustive. This second reason is likely in jurisdictions such as Australia and the UK, where citing every decision that considers the same question is cautioned against~\cite{Pd16of2013}.

Finally, while Cranfield style collections were, and still are undoubtedly important in empirical analysis of the effectiveness of a retrieval system, the lack of large collections for case law retrieval may only pose a problem academically and not for commercial search providers. Large commercial legal search providers can gain vast amounts of user data, with huge amounts of queries being issued~\cite{Shankar2018}. Vast query logs may mean that serving similar cases to users with similar queries may take precedence to optimizing effectiveness measures in a Cranfield style evaluation. This has definitely been an approach taken by Westlaw, where query logs are employed to generate surrogate documents from which features for a ranking functions can be learned~\cite{Liao2009}. 

Nonetheless, Cranfield style evaluations are still important, both where user data is not available, to ensure that a potential ranking method does not degrade search effectiveness too detrimentally, and for research, where in our view user-logs are not possible or limited largely. We view user-logs as infeasible in a research setting for several reasons. This relates to the sensitivity and confidentiality of legal searches. They are specific to advice to be given to, or cases being prepared for, a client. As such, without clients' consent, queries should not be contained in published query logs used for research. On the other hand, traditional Cranfield style collections can be more appropriately created. This is because, topics or queries can be expressed in an abstract fashion, without need for explicit approval from a client; a query or topic in and of itself does not give away anything about a client or their case and as such there is no issue of confidentiality. The impact of limited user data may be minimal however, given the short time required to obtain large amounts of data, and also where methods to generate synthetic user data are used~\cite{Pompili2019}. Creating such collections will undoubtedly be very costly. Cases are typically long and legal issues are complex. As such, in our view, lawyers will be necessary to assess relevance, as is done by commercial providers, who have in house teams of lawyers that perform these tasks.

While not specifically on the task of ad-hoc retrieval, several collections exist for closely related tasks. More recent COLIEE competitions include collections for finding cases relevant to a seed case, as well as identifying entailment between cited cases (how a cited case is treated)~\cite{Kano2018, Rabelo2019}.

\begin{table}[ht]
\centering
\begin{tabular}{p{.14\textwidth}p{.64\textwidth}p{.22\textwidth}}
\hline
\textbf{Source} & \textbf{Query} & \textbf{Generation Method} \\
\hline
Koniaris et al.~\cite{Koniaris_Multi_Dimension} & Products Liability & Topic in Westlaw Laws of America Digest \\
Turtle~\cite{Turtle_1994} & (741 +3 824) FACTOR ELEMENT STATUS FACT /P VESSEL SHIP BOAT /p (46 +3 688) ``JONES ACT'' /P INJUR! /S SEAMAN CREWMAN WORKER & Manually created by expert searcher\\
Turtle~\cite{Turtle_1994} & What factors are important in determining what constitutes a vessel for purposes of determining liability of a vessel owner for injuries to a seaman under the Jones Act (46 USC 688)? & Natural language issue statement \\
Locke et al.~\cite{AIRS} & ``sovereign immunity'' AND (immunity OR indemnif!) AND state AND suit AND (surrend! OR exist!) AND (tribe OR tribal OR ``indian trib!'') & Manually created by expert searcher\\
\end{tabular}
\caption{Locke and Zuccon's~\cite{SigirCollection} summary of queries previously employed in legal information retrieval studies. Koniaris et al.'s~\cite{Koniaris_Multi_Dimension} queries were artificially created from Westlaw Digest topics. Turtle's~\cite{Turtle_1994} queries were created by lawyers. "/P" is a proximity based Boolean "and" operator where the operands must occur within the same paragraph. Natural language queries from Turtle's work are included for comparison.\vspace{-15pt}}
\label{tab:queries}
\end{table}

%% file: literature-review/methods.tex
\section{Methods for case law retrieval}
\label{sec:methods}

\begin{table}[ht]
\begin{tabular}{lll}
\textbf{Method}                                     & \textbf{Section}                   & \textbf{Main works examined}                                      \\
\midrule
Boolean and natural language               & \ref{sec:boolean}         & \cite{Dabney_1986, Turtle_1994, Gerson_1999}    \\
Conceptual search and case-based retrieval & \ref{sec:conceptual}      & \cite{Matthijssen_1995, Uijttenbroek_2008}       \\
Question answering                         & \ref{sec:question}        & \cite{Penas_2010, Do_2017}                      \\
Query expansion                            & \ref{sec:query-expansion}       & \cite{Custis_2008, Sartori_2010, Boonchom_2010} \\
Query reduction                            & \ref{sec:reduction}       & \cite{AIRS, FireNotes, SigirCollection}         \\
Search Result Diversification                            & \ref{sec:diversification} & \cite{Koniaris_Eval_Diversification}         \\
Use of citation networks                   & \ref{sec:networks}        & \cite{Wiggers2019, Raghav_2016}                 \\
Deeper understanding of texts              & \ref{sec:semantic}        & \cite{Grabmair_2015, Nejadgholi_2017, Shankar2019, Buddarapu2019}
\end{tabular}
\caption{Summary of methods and main works examined in this survey that comprise these methods. \vspace{-25pt}}
\label{tab:summary-methods}
\end{table}

In Table~\ref{tab:summary-methods}, we summarise the main methods for case law retrieval as well as the main works for each method. As briefly touched on above, several methods have, in the past, been identified as the standard for case law retrieval. In the early 2000s, Moens~\cite{Moens_2001} identified some form of concept based retrieval, relying on manual indexing of terms as commonplace, with three models as being standard for case law retrieval: Boolean retrieval, vector space retrieval, and probabilistic retrieval. She reported that legal text retrieval is primarily based on concepts identified by domain experts rather than on full text retrieval, in which indexing was frequently manually done and hence is very time consuming and expensive. While indexing of terms may be automated, she identified several reasons for why this leads to effectiveness of retrieval systems that are not comparable to systems where terms are manually indexed. Maxwell and Schafer~\cite{Maxwell_2008} stated that there are two main approaches to legal information retrieval: (i) knowledge engineering; and (ii) natural language processing techniques. 

There has been no comparison of these methods for case law retrieval. In fact, much of the research comparing, empirically, the effectiveness of case law retrieval systems as summarised below deals with evaluating commercial search systems. This is a significant gap in the literature. Attempts have been made to address this deficiency in published research in the AI and law community. Conrad and Zeleznikow~\cite{ConradZeleznikow2013, ConradZeleznikow2015} have summarised the presence of evaluation in works in ICAIL. In ICAIL, the majority of presented works were theoretical. Despite this, works presenting algorithms are more frequently being evaluated empirically. While the lack of empirical evaluation of many methods discussed in this survey is problematic, these methods are still of interest. Not least for the intuitions and ideas behind their development. All that is further required is a more grounded analysis such as comparisons with baselines, human evaluation and statistical significance testing~\cite{ConradZeleznikow2015}.

A common starting point in any discussion of case law retrieval is the study by Blair and Maron~\cite{Blair_Maron}. But this is not a study considering case law retrieval -- it considered retrieval of documents in a discovery setting. Similarities between other legal search tasks may nonetheless be present. The study analysed the effectiveness of a commercial litigation support system --- an early precursor to technology assisted review for discovery. It involved the manual review of 40,000 documents. The study found that while users thought they had retrieved at least 75\% of relevant documents, they had retrieved close to 20\%. While this system did not evaluate retrieval of case law, it has been suggested that these results accurately reflect the effectiveness of case law retrieval systems \cite{Dabney_1986}. But then again, extrapolating these results to case law retrieval systems has been cautioned against~\cite{Austlii}. Blair and Maron hypothesised that the poor recall of the system in question may have been a result of semantic differences in the texts. Whether this is a problem is not known. Other external\footnote{~By external, we mean that they are performed on commercial systems by persons not employed by the organisation.} studies of the effectiveness of commercial systems are summarised in Table~\ref{tab:baseline-comparison}.
\begin{table}[t!]
\centering
\begin{tabular}{lrrrrrr}
\toprule
\textbf{Review}                    & \textbf{System}  & \textbf{Year}   & \textbf{Query type} & \textbf{Recall}    & \textbf{Precision}   & \textbf{MR} \\
\midrule
Dabney~\cite{Dabney_1986} & Westlaw    & 1986 &      Boolean      & 0.197      & 0.269        &     \\
                          & LexisNexis &     & Boolean      & 0.114      & 0.261        &     \\
\midrule
Dabney~\cite{Dabney_1993} & Westlaw    &     1993 & Boolean      & 0.322     & 0.124        &     \\
                          & LexisNexis &     & Boolean      & 0.264     & 0.115        &     \\
Gerson~\cite{Gerson_1999} & Westlaw    & 1999 &     Natural language      & \multicolumn{2}{c}{0.31} & 2.3 \\
                          & LexisNexis &    &  Natural language      & \multicolumn{2}{c}{0.37} & 2.5 \\
\midrule
Mason~\cite{Mason_2006}   & Westlaw    &   2006 &    Unknown      & -         & 0.810        &     \\
                          & LexisNexis &    &  Unknown      & -         & 0.740         &     \\
\end{tabular}
\caption{Evaluations of the effectiveness of commercial search systems for ad-hoc case law retrieval. Note that each work used different collections of documents, queries and relevance assessments. The only evaluation that reports the particular version of search system used is Gerson, using WIN and FREESTYLE respectively. \vspace{-20pt}}
\label{tab:baseline-comparison}
\end{table}

Dabney's~\cite{Dabney_1986} analysis has little detail by way of the size of the collections and the queries used for evaluation. His subsequent work~\cite{Dabney_1993}, however, detailed use of 23 areas of law summarised by practitioners (including a comprehensive set of relevant cases for that area of law). While the size of one collection is not known, the other contained over 1 million documents. Gerson~\cite{Gerson_1999} evaluated the effectiveness of the systems using 22 queries, measuring the greater of precision or recall at 20 documents. Mason~\cite{Mason_2006} evaluated the effectiveness of the two systems using the same collection of documents. The collection's size is not detailed. Mason judged the first 10 results from 50 queries according to the measures adopted by TREC. The way in which Mason~\cite{Mason_2006} determined relevance of a judgment to a query differs from the way adopted by Gerson~\cite{Gerson_1999} and Dabney~\cite{Dabney_1993}, who both used annotated areas of law: a legal topic and list of relevant cases as determined by lawyers as an exhaustive list of relevant documents. Mason~\cite{Mason_2006}, on the other hand, adopted a subjective determination of the relevance of the first 10 retrieved documents, classifying returned documents as relevant or not-relevant.

In earlier work Dabney~\cite{Dabney_1986} suggested four potential ways in which the effectiveness of a system could be improved, including: (i) the ability of a system to search for plurals or lesser common synonyms (he also noted, without detail, that Lexis and Westlaw both search for plurals); (ii) the provision of more help to a user in "choosing search elements"; (iii) moving away from strict Boolean rules for determining whether a document is relevant or not; and (iv) usage of citations.

The substantial differences in the recall and precision in these previous studies may be due to increases in the effectiveness across these commercial search systems. Or it may be put down to the lack of control as regards the queries, collection, relevance assessments and methods used to evaluate these systems in these studies.

Another comparison of methods to note is that of the FIRE IrLed task~\cite{Kulkarni_2017}. This task involved retrieval of cited decisions from prior decisions where the citation name had been removed. While one will immediately note that this is not the exact task of ad-hoc case law retrieval, it does provide some examination of methods. There were 200 cases for which citations were removed, and 2000 prior cases from which these citations were to be identified. The problem may be viewed as an identification problem: finding the correct citation for each occurrence within a document. However, the task became a ranking problem by virtue of the measures used, these being $MAP$, $P@10$, $Recall@100$ and $MRR$. A range of methods were considered. These included query reduction using statistical based term selection~\cite{FireNotes}, a language model using Dirichlet prior smoothing, probabilistic searching using BM25, a vector space model~\cite{Tian_2017}, ranking documents by the sum of scores from identifying the citation of legislative texts through regular expressions and returning prior decisions that also feature the same citations of articles, word distribution from an LDA topic model, and Doc2Vec~\cite{Le2014} cosine similarity~\cite{Kulkarni_2017}. Similar to the IrLed tasks, is FIREs AILA track~\cite{AILA}. One of the AILA tasks is finding relevant cases given a factual scenario expressed in natural language. Again, this task is slightly different to that of ad-hoc case law retrieval. Bhattacharya et al.~\cite{AILA} summarise the methods attempted in the AILA track.

\subsection{Boolean and natural language systems}
\label{sec:boolean}
\begin{table}[]
\resizebox{\textwidth}{!}{%
\begin{tabular}{lp{5cm}p{5cm}p{5cm}}
\textbf{Work} & \textbf{Summary} & \textbf{Evaluation} & \textbf{Findings} \\
\midrule
\cite{Turtle_1994}       & Comparison of Boolean and natural language search                & Cranfield style but non public collections                             & Natural language outperforms Boolean search                            \\
\cite{Dabney_1986, Dabney_1993} & Comparison of Boolean queries on two commercial systems          & Non-typical evaluation on commercial system using small number of queries & LexisNexis outperforms Westlaw in Recall and Precision                 \\
\cite{Mason_2006}        & Comparison of natural language queries on two commercial systems & Non-typical evaluation on commercial system using small number of queries & Westlaw outperforms LexisNexis in Precision                            \\
\cite{Gerson_1999}       & Comparison of natural language queries on two commercial systems & Non-typical evaluation on commercial system using small number of queries & LexisNexis outperforms Westlaw for combination of Recall and Precision \\
\cite{Maxwell_2008} & Summarises knowledge-base and natural language case law search & N/A & N/A \\

\end{tabular}%
}
\caption{Summary of works comparing Boolean and natural language search. \vspace{-20pt}}
\label{tab:boolean-summary}
\end{table}

Boolean retrieval represents the early foundation of case law retrieval systems. We summarise works that have investigated Boolean systems in Table~\ref{tab:boolean-summary}.

In our view, the most comprehensive evaluation of the effectiveness of Boolean methods for case law retrieval is that by Turtle~\cite{Turtle_1994}. Turtle, then a researcher at Westlaw, compared the effectiveness of Boolean and natural language queries on two collections of 12,000 and 410,000 documents,\footnote{~These collections were data internally available at Westlaw, and are used in their commercial system. We do not summarise these collections above, however the smaller is notable for its complete relevance assessements. } using Westlaw's case law retrieval system. This was an implementation of Turtle's inference network retrieval system~\cite{Turtle1991, TurtleThesis}. Evaluating the system's effectiveness over 44 queries, he concluded that evaluation on a commercial system shows natural language queries offered superior effectiveness to Boolean queries. On the smaller collection, his evaluation found a precision at 20 (P@20) and a recall of 0.423 and 0.244 for Boolean queries, and 0.57 and 0.329, respectively, for natural language queries. On the larger collection, his evaluation found a P@20 and a recall of 0.611 and 0.217 for Boolean queries and 0.759 and 0.269 for natural language queries. The evaluation, at what may be suggested to be a low depth for a recall orientated task - P@20, takes its motivation from Turtle's suggestion that lawyers are not willing to search through all results, but will evaluate only the first 20 or so before choosing to reformulate their query.

This remains the best, and in essence, the only study in the domain (aside from those discussed above that performed studies external to the system) as to both the comparative effectiveness of Boolean retrieval systems, and also of natural language systems. One criticism that may be made of the work is the method used to create Boolean queries compared to the natural language query. The natural language queries  were fixed. On the other hand, the searchers were allowed as much time to formulate the Boolean queries as they felt necessary. Despite this, these queries proved less effective.

Locke et al.~\cite{AIRS} have previously discussed Turtle's~\cite{Turtle_1995} criticisms of Boolean systems for, mainly, the lack of relevance ranking and result sets that are larger than what a user would be prepared to browse. This was such, as Maxwell and Schafer~\cite{Maxwell_2008} and Turtle had noted~\cite{Turtle_1994, Turtle_1995}, that the problem with Boolean retrieval is "that the larger the collection searched on, the greater the difficulty in achieving" higher precision~\cite{AIRS}. Schweighofer and Geist~\cite{Schweighofer_Geist}, in considering the need for query expansion, shared this view. As Locke et al.~\cite{AIRS} summarized, they noted that the effectiveness of Boolean queries might not be poor in the legal domain because "lawyers as domain experts will have knowledge of synonyms, without which effectiveness may suffer ... [but] domain knowledge has its limits, and one cannot reasonably know all other possible choices for a word"~\cite{AIRS}. This problem is common to both Boolean queries and best match retrieval. However, there is no empirical evidence to suggest this. 

This brings about questions of whether explicit methods (query suggestion) or implicit methods (query expansion) can address this potential problem. We discuss these methods for case law retrieval later. Scheweighofer and Geist~\cite{Schweighofer_Geist} argued that usage of term frequencies in the legal domain is not as helpful as other domains because ``no redundancy exists in legal norms, but a lot of information is irrelevant in case law. Relevant texts parts may consist only of a short paragraph or even only of a single sentence in a very long legal document.'' Kumar~\cite{Kumar_2014} discussed problems posed by the language used in legal documents. But, while not specific to case law, the problems are pertinent. Given the particular legal sublanguage, Kumar states that the statistical characteristics of words may be different from those in other more general corpora, with key terms appearing few times, and that there is no factorisation of the context in which such words may appear. Moens~\cite{Moens_2001} stated that legal language is not a pure sublanguage. Rather, it consists of a diverse vocabulary usually employing domain specific concepts. Further, it is unique in so far as legal language is typically composed of ``exceptionally long sentences and in crucial subclauses''~\cite{Moens_2001}.

In the context of web search, Carpineto et al.~\cite{Carpineto_2012} noted in 2009 that the average query length was 2.3 words, and that this average had not increased in the 10 years prior. As a result of the shortness of queries employed by information-retrieval (IR) users, the vocabulary problem has become ``even more serious'', and that the size of data makes ``polysemy more severe''. More recent query logs have shown queries to typically be of a similar length (an average of 2.4 words)~\cite{Xiong2019}. While web queries may be poor, there is no qualitative analysis to report on the effectiveness of queries in case law search. The only literature specific to case law retrieval is a comment by Shankar and Buddarapu~\cite{Shankar2018}, in a study of query intent classification, that queries commonly used in a commercial case law search system (LexisNexis) are very short. This leads to a poor ability to specify a user's information need. But it must be borne of mind that query length is task specific. In this regard, we note that as more question answering systems are included in commercial search systems, average query lengths may increase. This will be a result of more queries expressing information needs as questions. 

Finally, the length of the result set is not a problem unique to Boolean retrieval. As previously stated, the task of case law retrieval is viewed as a recall orientated task. But it is not a total recall task. Accordingly, in favouring recall over precision, with lawyers likely unwilling to wade through all results returned, ranking and explainability of results returned by a system are key. Early Boolean retrieval models ranked documents by date, and ranked retrieval by other means was not, at this time, implemented in commercial systems. The advantage to such a system was that it was easy to explain why a document is retrieved by a query, whereas it is much more difficult to explain ranking models despite the superior effectiveness that they offered. Explainability of a search system is important for reproducibility. But, as can be contrasted with, for example, systematic reviews where there is a strict protocol for searching, or technology assisted review where there are discussions regarding query formulation, this task is interactive and reproducibility is not strictly necessary. Reproducibility is even less so important as regards the question of a lawyer's searching being competent or negligent given search history logs. The more important question is, in our view, one of user satisfaction as to why a query returns certain results. One more point may be made of reproducibility. Near constant adding of cases and updating the treatment of cases may mean that results are not reproducible.



\subsection{Conceptual search, ontologies, case-based retrieval and argument retrieval}
\label{sec:conceptual}

Much research has focused on ontological or conceptual methods for case law retrieval. Most of this research focuses, however, on the creation of systems for practical applications rather than the evaluation of a system's effectiveness, or on systems that match cases to a set of factual circumstances or a seed case. It is not our intention to summarise all of the research in this regard.

The problem with simple full text matching, and therefore the need for a deeper understanding of the text being searched on is said to come from, as Matthijssen~\cite{Matthijssen_1995} put it, the "conceptual gap". There is in essence, a difficulty in translating an information need into a query because it is difficult to know what information solves a legal problem prior to finding it~\cite{Belkin1982}. As Carvalho~\cite{Carvalho_2017} noted, in legal texts, concepts are often used in different ways when compared to common language. Matthijssen stated that queries are often too unspecific; the particular "information need is lost in the query formulation process". One such way to overcome this problem is enhancing browsing features. However this falls outside the scope of our survey, given our focus on ad-hoc query search. This problem can be thought of as similar to that of vocabulary mismatch, where a user may not express their information need in the same manner as that represented in the documents searched on~\cite{Metzler2007}. Such a problem is common in Boolean retrieval~\cite{ElJelali_2015}. However, given the particular legal sublanguage of case law, where concepts are expressed in similar terms, vocabulary mismatch may only pose a problem when first approaching a search task. Once a user has found one relevant document or piece of information, interactively reformatting a query to use the language of found information may alleviate this vocabulary mismatch problem.


Klein et al.~\cite{Klein_2006} described ontological based methods for retrieving similar cases to a seed case. This is not in the context of ad-hoc  search, but rather so as to advise parties to litigation of similar cases based on their factual situation. They index documents  by removing stop-words,  reducing all verbs to a particular form, and mapping documents onto a conceptual structure using an ontology. They then retrieve documents using thesaurus-based statistical retrieval to identify relevant words. From this, each document is given a ``concept fingerprint'', being the list of concepts identified in a document as well as scores denoting relevance of that document to a particular concept. Represented as a vector, queries are then matched to documents in a vector space. They do not achieve great effectiveness with manual or automatically created ``fingerprints''. Gifford~\cite{Gifford_2017} discussed a new commercial legal case law retrieval system designed for finding arguments. This is slightly distinct from the task of finding case law as the primary goal of the system is, as Gifford put it ``for discovering and presenting legal arguments rather than entire appellate cases.'' The system used hierarchies with an ability to search over any part of the hierarchy.





Matthijssen~\cite{Matthijssen_1998} detailed a task-based index structure for retrieval. The index is structured so that terms correspond to tasks, being "a group of activities and procedural steps that are directed towards a common goal". This, Matthijssen argued, simplified the gap between a lawyer expressing a particular information need and task, however it increased the gap between indexing documents and the task. To address the same problem but from the perspective of laypersons, Uijjtenbroek et al.~\cite{Uijttenbroek_2008} and Laarschot et al.~\cite{Laarschot_2005} described a system for assisting the general public with case law retrieval. They used an ontology to map terms that a layperson would use to an ontology used for indexing decisions according to their legal concepts. This is one way to overcome another main problem with ontological methods, being, retrieval is dependant on expert domain knowledge~\cite{ElJelali_2015}. Another related way to overcome this problem is, more relevant to ad-hoc query search, query expansion with ontologies.

A key issue with the use of ontologies for ad-hoc case law retrieval is that, as Maxwell and Schafer~\cite{Maxwell_2008} recognised, with the amount of case law decided, manually annotating cases according to ontologies is not feasible. This means reliance must be placed on automatic methods for determining a case's ontological classification. Such a problem was noted as early as 2001~\cite{Moens_2001}. For this reason, focus on other methods for full-text retrieval is necessary. In somewhat of an attempt to combat this, and of relevance to ad-hoc query search, Rissland and Daniels~\cite{Rissland_1996a} detailed the combination of case-based reasoning (CBR) systems to supplement IR systems. As they recognised, CBR systems are deeply structured but contain very few documents, whereas IR systems contain many documents but represent them at a shallow level (text only).  Their system used a list of output cases from a CBR analysis as input to the INQUERY IR system~\cite{Callan1992}, to provide relevance feedback on this set in order to create a query for searching over the whole collection of case law; the "use of relevance feedback, in effect, tells the IR component that the cases found through the CBR analysis are highly relevant and that the IR engine should retrieve more like them." Two limitations are evident. First, this system is limited to taking a seed case as input and does not address ad-hoc query based retrieval. Second, necessarily finding cases in the initial CBR step will be limited to the scope of the CBR collection. Such a problem is also recognised by El Jalali~\cite{ElJelali_2015}. More recent work detailed by Conrad and Al-Kofahi~\cite{ConradKofahi2017} discusses large scale argumentation mining tools for jury verdicts, employed at Westlaw (scenario based search). Their work enables lawyers to view and group case outcomes from 500,000 cases according to a number of features. Accordingly, the problems identified by Moens~\cite{Moens_2001} and Rissland and Daniels~\cite{Rissland_1996a} may pose less of a problem as more and more machine learning based approaches are adopted.



\subsection{Question Answering}
\label{sec:question}

Many works have considered question answering for legal information retrieval tasks. These consider case law retrieval in that commercial systems return snippets of cases to users. More recent works by Westlaw detail many of their question answering systems~\cite{Custis2019, McElvain2019}. As is the trend with these works, and information retrieval generally, they employ machine learning methods and leverage Westlaw's vast user query logs to train the systems~\cite{Pompili2019}. WestSearch PLUS~\cite{McElvain2019} is the latest iteration of this. It provides a natural language interface to over 22 million single sentence summaries of case law, and high-confidence candidate answers are returned along with ad-hoc results returned from Westlaw's search system. Similarly, Bennett et al.~\cite{Bennett2017} demonstrated a system employed at LexisNexis that automatically identifies parts of documents that are extracted and treated as answers to be presented to users. This system suggests potential questions to users in a query box via autocompletion and returns an answer card above ad-hoc search results. As these are commercial systems, little is detailed by way of evaluation.

There is also a large body of work that considers retrieval of civil code articles based on a sample question from the Japanese Bar exam in the COLIEE tasks~\cite{Kim2017, Do_2017}, as well as multilingual retrieval over European Union legislative materials, in the ResPubliQA challenge~\cite{Penas_2010}. These works are outside the scope of this article. However, their approaches are nonetheless informative. Learning to rank and deep learning models trained using various features are commonplace. The investigation of these methods is nonetheless applicable to ranking of ad-hoc case law retrieval as commercial case law retrieval systems employ learning to rank methods~\cite{Mart2019}. The task also accounts for paragraph retrieval, which is something that has not been considered in the literature but passage or sentence level retrieval is suggested as being important to users of commercial legal search engines~\cite{LuConrad2012}.


\subsection{Query expansion}
\label{sec:query-expansion}
The goal of query expansion is to introduce terms in an effort to address vocabulary-mismatch problems; a goal similar to that of conceptual search methods. In Table~\ref{tab:expansion-summary}, we summarise research on query expansion in the considered domain. Outside the domain and more generally, Carpineto et al.~\cite{Carpineto_2012} provided a survey of automatic query expansion for IR. There has been little work exploring query expansion for case law retrieval.

Custis and Al-Kofahi~\cite{Custis2007, Custis_2008} are the only works exploring query expansion not using ontologies. These works were conducted internally at Westlaw on their commercial systems and using the benefit of their extensive user logs and data. It is pertinent to note the use of query expansion in many major commercial legal search systems~\cite{Wiggers2019}. Such use appears common place. Earlier work of Custis and Al-Kofahi~\cite{Custis2007} automatically degraded queries by removing terms, introducing term mismatch, to better evaluate the effectiveness of their expansion methods. They compared these methods to BM25 and Query Likelihood baselines.  They do not detail their expansion methods other than to say one is BM25 with pseudo-relevance feedback, and the other is a "language modeling based retrieval engine that utilizes a subject-appropriate external corpus (i.e., legal or news) as a knowledge source". The reference to Berger and Lafferty~\cite{Berger_1999} vis. translation probabilities suggests it follows the same approach as their later work~\cite{Custis_2008}. Custis and Al-Kofahi's empirical evaluation found that implicit expansion in their proprietary system greatly helped effectiveness. However, they found BM25 with pseudo-relevance feedback was less effective than the BM25 baseline. Over larger collections that did not contain case law, BM25 with pseudo-relevance feedback typically outperformed their proprietary method when more term mismatch was introduced at deeper measures (Recall@1000, MAP) compared to lower measures (Precision@10). 

Custis and Al-Kofahi~\cite{Custis_2008} do not use explicit expansion, instead relying on the retrieval model used, being the probabilistic translation language model proposed by Berger and Lafferty~\cite{Berger_1999}. They consider three methods to compute the translation probabilities: (i) an estimate using co-occurence probabilities of words within a window of words surrounding a term; (ii) a PRF based measure that limits the translation component of Berger and Lafferty's model to the top 50 terms from the top 20 documents retrieved by a search; and (iii) log analysis with 39 million click through events to estimate probabilities. Their analysis involved removing query terms from a query and comparing relevant documents. They found that, over 20,000 cases and 335 queries, query expansion significantly improved both MAP and recall at 1000. The results show that simply estimating the translation probabilities using surrounding words achieves good results, even when compared to estimating probabilities using query logs. It would be interesting to see what effect however, is had at lower depths, i.e. P@20 or P@50 given the authors' view that case law retrieval is interactive and lawyers are more likely to stop searching after 20 documents and reformulate their query~\cite{Turtle_1994}. We also note that recent works in IR have proposed alternative methods for estimating $P(q|D)$, the effect of which in this domain may be interesting to investigate~\cite{Karimzadehgan2010, Karimzadehgan2012, Zuccon2015}.

It is worth mentioning that use of query logs for case law retrieval is becoming more frequent. Their use can be seen in training legal question-answering systems~\cite{Custis2019, Pompili2019} as well as in recommending cases to users based on similar user profiles~\cite{LuConrad2012} (another tool outside of ad-hoc querying that is of great use to lawyers once they have found a relevant case for a given information need). Learning-to-rank based approaches adopted at Westlaw~\cite{Mart2019} might lead one to hypothesise that query logs are also applied to ad-hoc ranking for case law retrieval too.

\begin{table}[t!]
	\resizebox{\textwidth}{!}{%
		\begin{tabular}{lp{5cm}p{5cm}p{5cm}}
			\textbf{Work} & \textbf{Summary} & \textbf{Evaluation} & \textbf{Findings} \\
			\midrule
			\cite{Custis2007}        & Degraded query by removing query terms and evaluation of implicit query expansion proprietary concept Westlaw search system and BM25 with pseudo-relevance feedback             & Evaluated over internal collection of 11,000 case law documents and 44 queries with total relevance assessments & Query expansion led to significantly less decrease in effectiveness of degraded queries, however BM25 with feedback was less effective than standalone BM25 \\
			\cite{Custis_2008}        & Comparison of Berger and Lafferty's translation language model on Westlaw search system             & Evaluated on internal collection of 20,000 documents and 335 queries with annotations by lawyers & Query expansion improved effectiveness compared to keyword queries \\
			\cite{Schweighofer_Geist} & Practical implementation knowledge base query expansion system for public use                       & No quantitative evaluation                                                                                           & Query expansion improves quality of search results                 \\
			\cite{Sartori_2010}       & Practical implementation using ontology based expansion for retrieval of legal audio-video media    & No quantitative evaluation                                                                                           & Query expansion was promising                                      \\
			\cite{Boonchom_2010}      & Evaluation of ontologies automatically expanded from a seed ontology for finding relevant sentences & Evaluated over 2 collections of a total of 1,303 sentences                                                          & No comparison to a baseline method without query expansion         \\
			&                                                                                                     &                                                                                                                      &
		\end{tabular}%
	}
	\caption{Summary of works evaluating query expansion. \vspace{-15pt}}
	\label{tab:expansion-summary}
\end{table}

Ontological framework based methods represent the bulk of, and the remaining extent of, the literature as regards query expansion. The study by Breuker et al.~\cite{Breuker2004} is one of the earliest suggestions of using ontology based query expansion for case law retrieval. They suggested expansion with both terms that are superclasses, and also terms that are subclasses, of terms specified by a user. Similarly, Sartori et al.~\cite{Sartori_2010} and Schweighofer and Geist~\cite{Schweighofer_Geist} have both proposed query expansion in the legal domain using ontological frameworks. Scheweighofer and Geist~\cite{Schweighofer_Geist}, do not report quantitative results, other than to say that the improvement is as expected, "quite good". Likewise, Sartori~\cite{Sartori_2010} does not provide any detail of improvements in performance. Other research into query expansion for case law retrieval also lacks quantitative evaluation~\cite{Subburaj_2016, Zaidi_2005, Breuker_2002}. Saravanan and Gavindran~\cite{Saravanan_2009} considered manual query expansion as well as suggestion through an interface that suggests terms based on an ontological framework. While they report greater performance in terms of precision and recall, they do not consider automated methods for expansion (users expansions are selected by users). Similarly, Tantisripreecha and Soonthornphisaj~\cite{Tantisripreecha_2009} considered query expansion by means of a domain ontology to improve result diversity. They create a set of queries where each contains an ontological concept, based on concepts related to the original query concept. They then issued these queries and rank each query individually, combining all results at the end to compile a ranked result list. They noted increases in diversity of results. Again however, their testing on two collections, each of less than 1000 cases, and just over 10 topics may mean that their results do not generalise.

Boonchom and Soonthornphisaj~\cite{Boonchom_2010} described a manual ontology based expansion system, where by traversing an ontology more terms are suggested to the user. The system also weighted ontological concepts. They reported precision over a small collection of 1300 sentences and thus their empirical findings may not generalise. This collection is split into two distinct topics, however the size of each is not known. On one collection, the weighted ontology obtained a precision of 0.89, whereas the non-weighted ontology obtained a precision of 0.72. Results on the other collection were lower for both weighted and unweighted approaches. Unfortunately their work does not compare the ontologies to a baseline method for retrieval of the sentences.

Finally and relatedly, Savelka et al~\cite{Savelka2019} attempted query expansion when searching for sentences in case law that discuss statutory provisions. They did this by two means: including words from the provision itself; and, taking similar words from word embeddings learned over the collection of cases to those words found in the query. But their evaluation found that expansion by selecting similar terms did not increase retrieval effectiveness.

\subsection{Query reduction}
\label{sec:reduction}
\begin{table}[]
\resizebox{\textwidth}{!}{%
\begin{tabular}{lp{5cm}p{5cm}p{5cm}}
\textbf{Work} & \textbf{Summary} & \textbf{Evaluation} & \textbf{Findings} \\
\midrule
\cite{AIRS}        & Reduction methods based on several term scoring methods              & Evaluated over collection of 60,000 cases and 100 queries & Query reduction improved effectivenss of retrieval system \\
\cite{SigirCollection}       & Reduction methods based on several term scoring methods              & Evaluated over collection of 3 million cases and 12 queries & Query reduction improved effectivenss of retrieval system \\
\cite{FireNotes} & Reduction methods based on several term scoring methods used to find cited cases &  Evaluated on collection of cases where cited cases were anonymised & Reduction is promising method to identify cited cases \\
\end{tabular}%
}
\vspace{-15pt}
\caption{Summary of works evaluating query reduction. \vspace{-20pt}}
\label{tab:reduction-summary}
\end{table}

Query reduction consists of selecting a subset of terms from a query to construct a more effective query~\cite{AIRS}. This is typically in the context of more verbose queries. As with other facets of case law retrieval, little work has examined query reduction.  We summarise these works in Table~\ref{tab:reduction-summary}: the only works that have examined the application of keyword extraction or query reduction methods are those of Locke et al.~\cite{AIRS} and Locke and Zuccon~\cite{FireNotes, SigirCollection}.

Locke et al.~\cite{AIRS} considered several measures for term scoring for automatic reduction of text from cases and compared these to the effectiveness of manually generated queries. They took a similar approach to that described in the work of Koopman et al.~\cite{koopman2017generating} with regard to generating queries for a range of proportions (Koopman et al. considered query reduction in the medical domain). To score terms for inclusion in a query, they evaluated IDF, parsimonious language model and Kullback-Leibler informativeness. Similar approaches were adopted in Locke and Zuccon~\cite{SigirCollection}, as well as in Locke and Zuccon~\cite{FireNotes}, but this was not in the context of ad-hoc query retrieval, rather in seeking to find cited cases within a case given the text surrounding a reference to a case. 

Research in the domain has only considered unigram based statistical methods for selection of appropriate terms when automatically generating queries. Logically therefore, methods that focus on the multi-word terms (n-grams) or syntactic phrases may be an appropriate next step for consideration of automatic query reduction methods.

\subsection{Search Result Diversification}
\label{sec:diversification}
\begin{table}[]
\resizebox{\textwidth}{!}{%
\begin{tabular}{lp{5cm}p{5cm}p{5cm}}
\textbf{Work} & \textbf{Summary} & \textbf{Evaluation} & \textbf{Findings} \\
\midrule
\cite{Koniaris_Eval_Diversification}        & Diversification using various methods              & Evaluated on collection of 60,000 cases with automatically generated relevance assessments & Search result diversification methods outperform text summarisation methods and \texttt{tf-idf} baseline \\
\end{tabular}%
}
\caption{Summary of works evaluating search result diversification. \vspace{-20pt}}
\label{tab:diversification-summary}
\end{table}

We summarise the works involving diversification in Table~\ref{tab:diversification-summary}. Diversification requires trading off finding relevant documents for a diverse set of documents for a given query~\cite{Santos2015}. It is important to note that this may not be an appropriate task given the suggested nature of case law retrieval as a recall orientated task, but one concerned with precision. If the task is one of finding relevant seed documents the time taken to complete a given task is of key importance. The ranking of results may differ for a diversified set, and this will lead to an effect on the time taken for a searcher to find all relevant decisions and to complete a search task. However, this is not something that can be easily quantified, nor has research attempted to do so. But, if case law retrieval is a total recall task, diversification, which will only affect the ranking of documents and not the retrieved set, will merely only affect user satisfaction with a given system. 

Koniaris et al.~\cite{Koniaris_Eval_Diversification} stated that "it is extremely difficult to search for relevant [cases] by using Boolean queries", and that diverse results are "intuitively" more informative than homogenous result sets that contain results with similar features. They concluded that diversification methods "demonstrate notable improvements in terms of enriching search results with other hidden aspects of the legal query space", and after empirically evaluating the methods according to a-nDCG (alpha-normalised discounted cumulative gain), ERR-IA (expected reciprocal rank intent aware) and S-recall (subtopic recall), web search diversification methods were the best methods evaluated. Koniaris et al.~\cite{Koniaris_Eval_Diversification} sought to return a set of 30 diversified results from an original set of 100 after an original ranking. They considered these results at cutoffs of 5, 10, 20 and 30 for varying levels of interpolation between relevance and diversity, following a ranking of results produced using cosine similarity and log-based TF-IDF methods. While various methods of diversification are considered, they evaluated their methods over a collection of 3890 cases, for which they used automatically obtained relevance assessments from an LDA topic model trained over the highest scored TF-IDF documents for a query. The results may not be generalisable, and may be different had manual relevance assessments been obtained.

%% file: literature-review/network-analysis.tex
\subsection{Use of citation networks}
\label{sec:networks}
Citations are an important part of a legal decision. Judges cite for a number of reasons. In most instances, there is an obligation to give reasons for a decision. For a judge to show that their decision conforms with established and binding principle, they must explain their conclusion. In so doing, they will state the law, by reference to a particular decision, or to another source of law. They will, broadly speaking, say why their decision follows a particular law, or does not. A citation by a judge of a previous decision may express a range of acceptance of the cited decision, from following the decision, to distinguishing (saying that it does not apply to the factual situation), to doubting its application, or rejecting it.

\begin{table}[t!]
	\resizebox{\textwidth}{!}{%
		\begin{tabular}{lp{5cm}p{5cm}p{5cm}}
			\textbf{Work} & \textbf{Summary} & \textbf{Evaluation} & \textbf{Findings} \\
			\midrule
			\cite{Tapper_1974}        & Earliest proposal for use of citations to rank case law by representing a case as a vector of citations &  Manual comparison of seed to highest ranked cases & Citation vectors provide a means for finding similar cases \\
			\cite{Wiggers2019}        & Comparison of citation networks in Dutch case law              &  Manual evaluation of citation types in case law &  Citations are likely a useful feature for ranking case law   \\
			\cite{Gelbart_1990, Gelbart_1991}        & Use of citations as feature in vector-space model in commercial system             &   Not considered &  Not considered  \\
			\cite{Koniaris_Eval_Diversification}        & Use of citations to diversify case law results             &   Cranfield style evaluation on automatically created collection & Citations provide effective diversification of case law search results  \\
		\end{tabular}%
	}
	\caption{Summary of works evaluating use of citation networks for case law retrieval. \vspace{-20pt}}
	\label{tab:citation-summary}
\end{table}

Citation analysis in case law has been frequently studied. For instance, Agnoloni et al.~\cite{Agnoloni_2014},  Fowler~\cite{Fowler_2007}, Neale~\cite{Neale_2013}, Koniaris et al.~\cite{Koniaris_Network}, Geist~\cite{Geist_2009} and van Opijnen~\cite{Opijnen_2012} present large studies of citation networks in case law. With regard to case law retrieval, citation indexes have existed for some time \cite{Zhang_Koppaka}.\footnote{ Two commercial services exist: LexisNexis Shepard and Westlaw Keycite. As do other freely available systems such as LawCite: \url{www.austlii.edu.au/lawcite}. } Works by persons at Westlaw have said that citations are used for ranking case law~\cite{Turtle_1995, ConradNextGen2013}, as well as their use in other search systems~\cite{Mart2019}, but, unsurprisingly, there has been little publicised application of network analysis methods towards improving effectiveness of case law retrieval systems. We summarise these works in Table~\ref{tab:citation-summary}.

The earliest proposal of the use of citations in case law retrieval is by Tapper~\cite{Tapper_1974}. First as a means of tracing relevant cases~\cite{Tapper_1974}. Several commercial systems provide tools for doing this, by visualising a decision's citation network.\footnote{ See \url{www.ravellaw.com} and \url{www.jade.io}.} Later, Tapper proposed citations as a means for ranking by representing cases as a vector of their citations, as opposed to a vector of its terms~\cite{Tapper_1981}. Despite: (i) the creation of visual tools for mapping citation networks; (ii) suggestions that citations are a useful ranking feature~\cite{Wiggers2019}; and (iii) findings that legal citation networks exhibit similarities as compared to other networks \cite{Geist_2009} that have been used to rank results of search systems, citation networks have only been applied to ranking documents for case law retrieval in two instances. This can be contrasted with the employment of such measures in web result ranking~\cite{Page_1999, Metzler2011}.

The FLEXICON system combined citations as part of the ranking process in a vector space model for searching over a very small collection of case law (1000 documents)~\cite{Gelbart_1990, Gelbart_1991}. As a commercial system, the exact way in which citations were used was not detailed.

Koniaris et al.~\cite{Koniaris_Eval_Diversification} considered an application of DivRank in seeking to diversify search results. They found it, and other web-based diversification methods were effective. 





While not in the context of reranking results for ad-hoc queries, Winkels et al.~\cite{Winkels_2014}, used network analysis to suggest decisions that are relevant to a particular legislative provision. This approach is similar to Tapper's early approach for finding similar cases to a seed case using citation vectors~\cite{Tapper_1981}, where documents were represented as a series of citation vectors rather than term vectors. This has one notable drawback: use of citation vectors only to rank cases rather than as a feature means that cases that are not cited, or do not cite other cases will not be ranked. Such an outcome is undesirable and may happen where a case considers a new, uncommon, point of law, or the case is newly decided and has not been cited by any other cases.

As Winkels et al.~\cite{Winkels_2011} noted, many of the most cited cases (in a network of Dutch cases) deal with procedural issues rather than substantive issues. Substantive laws are those which identify and define rights, whereas procedural laws deal only with regulating court proceedings: \textit{McKain v RW Miller \& Co (SA) Pty Ltd} (1991) 174 CLR 1.

This focus of cases on procedural issues is illustrative of one of the problems with the use of a network to rank documents. Panagis et al.~\cite{Panagis_2017} recognised the different circumstances in which a decision may be cited and that simple network analysis of citations falls short in this regard by equally treating citations; a citation may be made without much application of any principle or it may be made for a number of other reasons, from approving, following or dissenting from (expressing disagreement with), a decision. Unlike webs page where it may be safe to assume that the page deals only with one topic, case law frequently involves multiple topics. For instance, a case may decide any number of questions, each unrelated to the other. In this respect, topicality of the citation, or the manner in which the cited case is treated are interesting areas that have not yet been explored. Any consideration of citations for ranking should consider both the topicality of the citation and the treatment of the cited case. 

While not for ad-hoc search, Zhang and Koppaka~\cite{Zhang_Koppaka} from LexisNexis, proposed a system that analysed the citation network of a case based on the topic for which a case was cited. This system allowed users to follow links based on the legal topic. The system involved discerning the reasons for which a case was cited through looking at the citation's surrounding sentences. Similarly, commercial systems offer tools for tracing citations at a paragraph level.\footnote{ Jade Barnet offers such a tool: see \url{jade.io/j/?ht=FAQ+-+Focus+Matches&t=help}.}

In terms of such tools being incorporated into ad-hoc retrieval, Panagis et al.~\cite{Panagis_2017} explored citations at a paragraph level rather than at a document level, in combination with text analysis to determine non-explicit references. But, a limitation of their work is that they did not consider ranking. Raghav et al.~\cite{Raghav_2016} ranked case law based on a combination of two measures. The first measure split a seed judgment into a set of paragraphs and calculated BM25 similarity between the paragraphs of all other judgments. The second measure used was bibliographic coupling to calculate similarity between two judgments. Bibliographic coupling is the number of citations to another document that two documents share~\cite{Kessler1963}. This ranking was based on an existing seed judgment rather than a query, however. Given the above comments regarding topicality at a paragraph level, the use of bibliographic coupling is odd, given that this will calculate the similarity between judgments overall based on the similarity of their citations. More recently, more complex methods for identifying similarities between cases have been proposed. These take into account not just a computed similarity between the citations of two cases, but also the similarity between references to legislation~\cite{Bhattacharya2020}.

%% file: literature-review/rhetorical-roles.tex
\vspace{-8pt}
\subsection{Deeper understanding of legal text and AI applications to case law retrieval}
Applying AI methods to legal problems has been a topic of substantial interest in the past 20 or so years. The International Conference for Artificial Intelligence and Law shows a depth of work in this area. We do not wish to repeat much of the work that has occurred in this domain, as it is on other legal tasks and outside of the scope of this survey. However, for useful surveys of the works in ICAIL see Conrad and Zeleznikow~\cite{ConradZeleznikow2013, ConradZeleznikow2015}.

Shankar and Buddarapu~\cite{Shankar2019} detail character level machine translation models for query reformulation, word-embedding based query-intent understanding~\cite{Buddarapu2019}, as well deep ensemble learning query understanding techniques~\cite{Shankar2018} used at LexisNexis. Query reformulation and understanding all form important parts of the ad-hoc retrieval process, ensuring queries are routed to a correct vertical search engine and any entities in a query are correctly identified. As Shankar and Buddarapu recognise, this represents a step away from traditional search and towards predictive coding. Other applications of AI to law by Westlaw and LexisNexis include automatic detection of overruling of cases and parsing cases for use in litigation analytics tools~\cite{Custis2019}.

Semantic search for ad-hoc case law retrieval has begun to be investigated. Relatedly, see the above discussion on query expansion based on translation probabilities. Given the move towards predictive coding, this is an area that warrants further investigation. Only one work has considered ranking case law using similarities between embeddings for sentences or paragraphs of a case~\cite{Sarsa2019}. Sarsa and Hyvonen~\cite{Sarsa2019} described a prototype system that ranked Finnish case law given a seed document using, amongst other methods, Doc2Vec~\cite{Le2014}. This is an analog to systems provided by commercial search providers, that combine numerous features including recommending based on similar user profiles~\cite{LuConrad2012}. Savelka et al~\cite{Savelka2019} used similarities between embeddings to find sentences in case law that are relevant to statutory provisions. Landthaler et al.~\cite{Landthaler2016} used word-embeddings to find similar legal provisions in documents and in statutes. Word embedding similarities have also seen use in other legal tasks such as e-Discovery~\cite{VoPrivault2017}. No work has specifically considered leveraging semantic similarities for ranking for ad-hoc case law retrieval. 

\label{sec:semantic}
\begin{table}[]
\resizebox{\textwidth}{!}{%
\begin{tabular}{lp{5cm}p{5cm}p{5cm}}
\textbf{Work} & \textbf{Summary} & \textbf{Evaluation} & \textbf{Findings} \\
\midrule
\cite{Grabmair_2015}        & Annotated portions of a case for use refining space of document text searched on              & Non-Cranfield style evaluation over collection of 188 cases & Poor classification of sentences detrimentally affected retrieval performance  \\
\cite{Nejadgholi_2017}  & Annotated sentences to train classifier to then search for sentences relevant to a query & Non-Cranfield style evaluation & Annotating sentences is effective means for finding factually similar sentences \\
\end{tabular}%
}
\caption{Summary of works evaluating semantic roles of text. \vspace{-30pt}}
\label{tab:semantic-summary}
\end{table}

Closely related, two works have considered the impact of classification of text according to its rhetorical role, as a means of retrieving relevant decisions~\cite{Grabmair_2015, Nejadgholi_2017}. We summarise these works in Table~\ref{tab:semantic-summary}. We note that similar applications in the context of literature retrieval exist~\cite{Boudin_2010, Scells_2017}. 

Semantic roles of text refers to what role the text plays within the document: i.e. is it conclusive or factual. In case law this will be whether the text is factual or legal. As the doctrine of precedent requires that like circumstances be considered in a similar fashion, being able to search a collection solely for similar factual situations may prove useful to lawyers. Similarly, lawyers as domain experts are often required to apply reasoning from other legal areas. As a result, the ability of a lawyer to search for law that is logically similar but not factually relevant may also prove useful. As an example, the meaning of words are often key to interpreting a piece of statute -- if the statute prohibits something that \textit{may} give rise to another thing, the meaning of \textit{may} could be determined by reference to legal decisions considering the word. And, the ability to search for legal statements that explain the meaning of the word would therefore prove useful. The intuition is therefore that searching on a confined part of the whole collection may lead to both precision and recall. 


The only study that has considered such an approach while ranking results is that of Grabmair et al.~\cite{Grabmair_2015}, who evaluated retrieval in the context of searching through a collection of 188 annotated vaccine related adverse reaction legal cases. They considered annotation of rhetorical roles for the purpose of ranking. They retrieved all sentences relevant to a query and then ranked documents based on the number of relevant sentences a document contains. Again to be noted is the small collection on which evaluation is undertaken, a common problem with case law retrieval evaluations.

The only other work is that of Nejadghole et al.~\cite{Nejadgholi_2017}, who implemented a semantic search system that searched only on facts of cases. They did so by training a classifier on a small number of annotated cases, and a word embedding model to automatically annotate a collection. They manually annotated 150 decisions out of a collection of 46,000 decisions, for a total of 12,220 sentences. Their annotations classified 46\% of the sentences in these judgments as fact. From their manual annotations, they trained a binary classifier to automatically annotate the remaining decisions in their collection. In order to evaluate retrieval performance, they created 15 queries, combined with 5 sentences for each query to compare whether they were in fact relevant or not relevant. They ordered results according to similarity to the query. They reported an $MAP$ of 0.78. While this research reports an interesting application of methods applied outside of legal IR to the retrieval of case law, it does not do so in a way that uses such a method for ranking in ad-hoc querying. By way of comparison, there have been several applications of rhetorical roles to ranking in the task of retrieval of systematic reviews~\cite{Boudin_2010, Scells_2017}.

There are obvious difficulties in retrieval based on this approach. Moens~\cite{Moens_2001} identified that low density training examples pose a problem for text categorisation. This is ultimately the problem that Grabmair et al~\cite{Grabmair_2015} faced, given that ranking ignoring a sentence's classification resulted in better performance. In related work in another professional search field, Boudin~\cite{Boudin_2010} stated that the performance of a weighted field approach may not work well because of difficulty arriving "at a consistent tagging of [the] elements.'' Such a difficulty is nonetheless present in annotating case law. But, there would appear to be large amounts of work in the automatic summarisation field that could be leveraged~\cite{Saravanan_2006, Shulayeva_2016, Grabmair_2015, Hachey_Grover}. Again, we note that this is extremely similar to tools such as Westlaw's scenario search, the difference being the use of natural language querying as opposed to searching through filtering different fields~\cite{ConradKofahi2017}. As we noted above, the use of machine learning techniques appears to have done away with much of the problem identified by Moens~\cite{Moens_2001}.

%% file: literature-review/query-expansion.tex
%
%

%% file: open-issues.tex
\section{Open issues, challenges and directions} 
\label{sec:open-issues}
Improving the effectiveness of ad-hoc case law search still faces fundamental challenges. Despite the size of the legal research industry and the legal industry that relies on it, little has been published on ranking for ad-hoc retrieval in the last 20 years. Further, there is no standard collection for evaluating methods, and this results in the difficulty in comparing any potential methods. The main focus of work in legal IR is on the semantic understanding of case law text, or legal text in general, or technology-assisted review. This work on semantic understanding is promising, but tends to be performed on a much smaller scale compared to typical IR evaluations: hundreds of documents instead of hundreds of thousands or millions of documents, with the exceptions of works by commercial legal search providers. Similarly, evaluations are not typical Cranfield style evaluations, but as we identified above, this is only a problem in an academic settings, and not commercially. Both of these problems mean that the conclusions reached by any evaluations may not hold true on larger collections or in real world situations. 

Intrinsic characteristics of case law documents make problems very difficult in this domain. As discussed in Section~\ref{sec:history}, the extreme length, number of diverse topics in a single document, complex language and structure of the texts' pose unique difficulties not often present in other domains. Likewise, debate about whether the task is one requiring total recall or, whether as in the authors' view, it is one requiring high recall but still requiring acceptable levels of precision introduces different problems. If the system is one requiring total recall replicability, explainability and user trust are paramount~\cite{RussellRose2018}. Ranking is less important. If the task is precision orientated, finding relevant documents within a small amount of interactions and therefore ranking is more important. 

As to appropriate future directions of research, outside establishing baselines and creation of a robust test collection on which to evaluate the effectiveness of methods, several areas warrant exploration. Firstly, the impact of domain specific features on ranking in case law retrieval: citations, which are prominent in case law, topicality, treatment and the relationship between a citing court and cited court, as well as the time between the cited decision are all matters not considered in the literature. Any such exploration of these features has not empirically evaluated their use for case law retrieval. Secondly, semantic representations of text are a typical focus point in many areas of research. This has focused on ontological representations of text, or QA systems, all of which have been on a small scale, with the exception of recent works by commercial search providers. Consideration of semantic search, or segmenting texts according to the role a subsection plays within an overall document all represent interesting directions for research.

%% file: conclusion.tex
\section{Conclusion}
\label{sec:conclusion}
Case law is judge made law that forms one of the key areas of law in common law legal systems. Case law retrieval systems are a key tool for lawyers to be able to effectively and cost efficiently carry out their work. Despite large commercial systems existing for a number of years, little research compares the effectiveness of methods for ad-hoc query retrieval of case law. This survey has summarised the extent of literature in this regard.

In light of the existing literature, several key observations may be made. In the last 30 years, the literature has reported a turn towards natural language queries. This is an incident of Boolean retrieval becoming ever less common~\cite{Bintliff2013, Poje_2014} and lack of teaching in legal education will only further this. However, Boolean searches, while less frequent than natural language searches, are still an important aspect of case law search~\cite{Shankar2019}. More and more research involves the application of machine learning techniques to both information retrieval and to legal tasks generally. Much of which has focused on conceptual search, and question answering.  Question answering systems are taking the forefront in commercial legal search systems. Little research has been published as to ranking for ad-hoc retrieval, as opposed to finding similar judgments to seed cases. This is not surprising for several reasons: the wealth of research in other legal search areas such as e-Discovery and other assistive legal search tools; and further, given the size and competitiveness of the legal search market, commercial operators may be unwilling to disclose the factors behind any success. Query expansion is adopted by many commercial search systems but there exists little published research empirically evaluating the efficacy of such approaches. The lack of standardised testing over large collections is one of the biggest problems in ad-hoc case law retrieval and creating a collection will be a time consuming approach, that most likely must be done by lawyers. But this is not a problem for commercial search providers, who can leverage the use of vast query logs~\cite{LuConrad2012, McElvain2019}.

Undoubtedly, the future of case law retrieval will be natural language based as Boolean queries are taught less in legal education~\cite{Bintliff2013}, and question answering systems take the forefront in commercial search platforms. But, for effective research to take place, outside of commercial ventures, large scale collections are required, so too are baselines of other state-of-the-art retrieval methods from other IR domains.